**Femtosecond Synchronization of Optical Clocks Off of a Flying Quadcopter**

Hugo Bergeron[+], Laura C. Sinclair*[,+], William C. Swann, Isaac Khader, Kevin C. Cossel, Michael Cermak, Jean-Daniel Deschênes*, and Nathan R. Newbury

[1]*National Institute of Standards and Technology, 325 Broadway, Boulder, Colorado 80305*
[2]*Université Laval, 2325 Rue de l'Université, Québec, QC, G1V 0A6, Canada*
[+] These authors contributed equally to this work.

*laura.sinclair@nist.gov; octosigconsulting@gmail.com; nathan.newbury@nist.gov*

Future optical clock networks will require free-space optical time-frequency transfer between flying clocks. However, simple one-way or standard two-way time transfer between flying clocks will completely break down because of the time-of-flight variations and Doppler shifts associated with the strongly time-varying optical link distances. Here, we demonstrate an advanced, comb-based optical two-way time-frequency transfer that can successfully synchronize the optical timescales at two sites connected via a time-varying turbulent air path. The link between the two sites is established using either a quadcopter-mounted retroreflector or a swept delay line at speeds up to 24 m/s. Despite 50-ps breakdown in time-of-flight reciprocity, the sites' timescales are synchronized to < 1 fs in time deviation. The corresponding sites' frequencies agree to ~ $10^{-18}$ despite $10^{-7}$ Doppler shifts. This work demonstrates comb-based O-TWTFT can enable free-space optical networks between airborne or satellite-borne optical clocks for precision navigation, timing and probes of fundamental science.

Optical clock networks promise advances in global navigation, time distribution, coherent sensing, relativity experiments, dark matter searches and other areas[1–12]. Such networks will need to compare and synchronize clocks over free-space optical links between moving airborne or satellite-borne clocks. However, current comb-based optical two-way time-frequency transfer (O-TWTFT)[13–15] cannot support femtosecond clock synchronization in the presence of motion. Even modest closing velocities between clocks lead to many picoseconds of non-reciprocity in the two-way optical time-of-flight and correspondingly large time synchronization errors. Here, we demonstrate an advanced comb-based O-TWTFT to synchronize clocks without penalty despite strong effective closing velocities. We synchronize two optical timescales connected via a quadcopter-mounted retroreflector or swept delay line over turbulent air paths at speeds up to 24 m/s. The synchronized clocks agree to $\sim 10^{-18}$ in frequency, despite $10^{-7}$ Doppler shifts, and to $<1$ fs in time deviation, despite 50-ps breakdown in time-of-flight reciprocity.

There are multiple challenges in implementing sub-femtosecond time-frequency distribution between moving clocks via free-space optical links. These challenges reflect and extend those faced by rf/microwave time-frequency transfer over free-space[16,17] and optical time-frequency transfer via fiber optics[7–10,18–24]. First, because of turbulence and diffraction, the received free-space signals will be weak, vary strongly, and suffer frequent fades. These turbulence-induced effects are far less for rf links, because of the longer wavelength, or for fiber-optic links, because of the stable medium. Previous comb-based O-TWTFT has nevertheless overcome these turbulence effects to achieve femtosecond synchronization[13–15]. Here, we focus on the second critical challenge. Namely, the clock sites can move rapidly, leading to strong Doppler shifts and a complete breakdown in the reciprocity of the two-way time-of-flight. Consider even a terrestrial velocity of 30 m/s. The fractional Doppler shift of $10^{-7}$ must be suppressed by $10^{11}$ to synchronize clocks to $10^{-18}$ in frequency. At this same velocity, the non-reciprocal time-of-flight of 3 ps (due to the finite speed-of-flight) must be suppressed by $10^4$ to synchronize clocks to below 1 fs in time. This level of suppression is orders-of-magnitude beyond that achieved in rf/microwave time-frequency transfer. Moreover, it must be achieved despite recurrent turbulence-induced signal fades. Here, we demonstrate an advanced comb-based O-TWTFT that synchronizes clocks to within femtoseconds despite motion by rigorously accounting for relativistic and systematic timing effects.

**Results**

Advanced O-TWTFT System

We synchronize two sites A and B each with a clock, or optical timescale, defined by the labelled pulses from a 200-MHz fiber frequency comb phase-locked to a ~195-THz local optical oscillator. (For a full atomic clock, this optical oscillator would be locked to an atomic transition.) Site A acts as the "master site". Site B is synchronized to it by adjusting the phase of the site B frequency comb. Because a fully "flyable" optical clock/oscillator is currently unavailable, both sites are fixed and we instead change the distance between sites by bouncing the optical signals off a quadcopter-mounted retroflector or a rapidly swept delay line. In either case, the link also includes the 2- or 4-km free-space turbulent air path. As shown in Fig. 1, the link is folded to enable verification by a single short fiber link that directly connects the sites to provide out-of-loop verification of the time synchronization[13–15]. All O-TWTFT information traverses the 2-4 km open-path link as if the two clock sites were, in fact, separated by this distance.

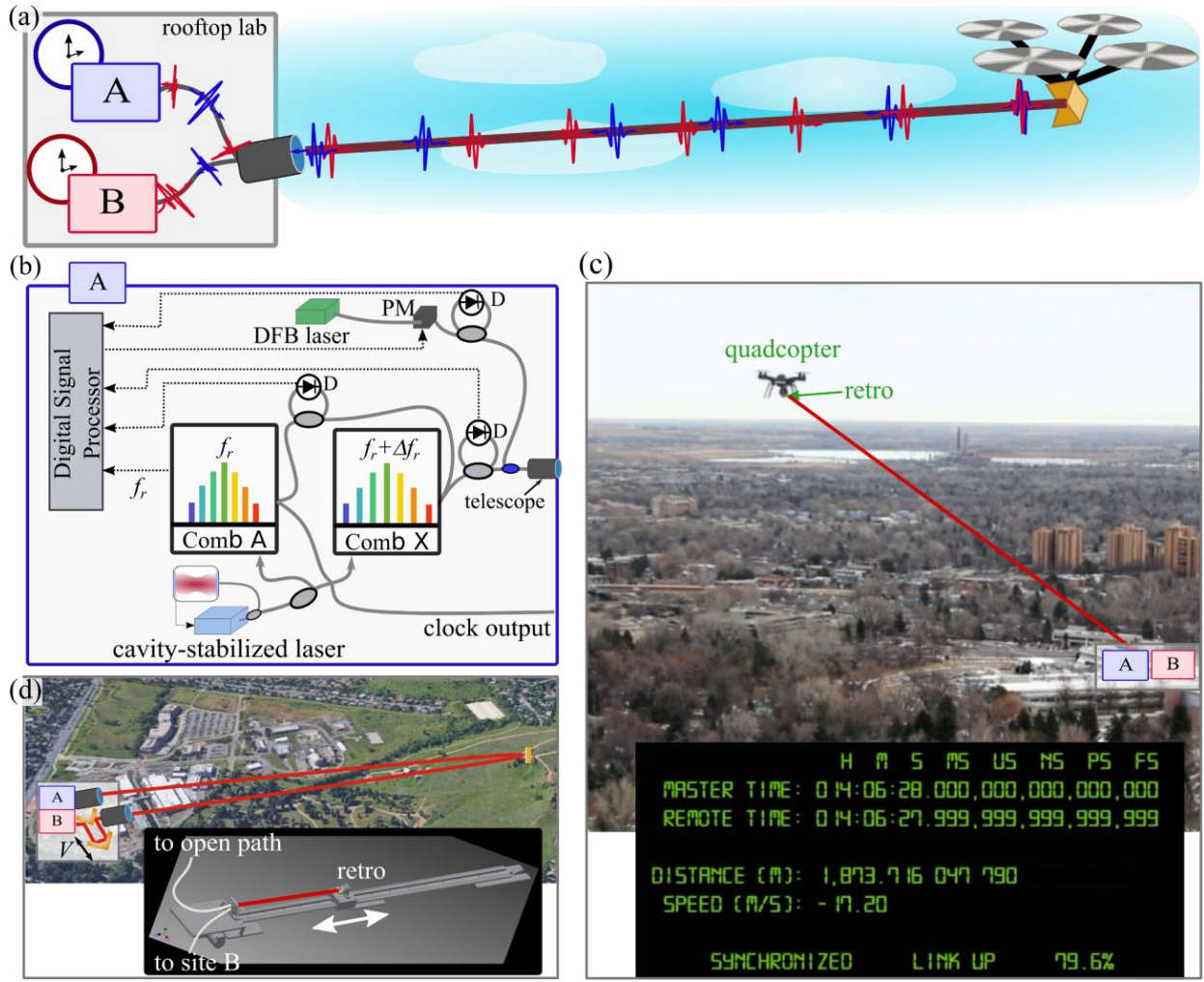

Figure 1. (a) Two optical timescales are synchronized over a folded link to a moving quadcopter-mounted retroreflector. The light is polarization multiplexed between site A and site B, co-located in a rooftop laboratory, and directed over the air to the quadcopter by a tracking telescope. (b) Schematic of master site A consisting of the comb A with repetition rate $f_r$, transfer comb X with repetition rate $f_r + \Delta f_r$, a cavity-stabilized laser, a phase-modulated distributed feedback (DFB) laser to support the optical communication channel and a digital signal processor. Grey lines: optical fiber; grey ovals: 50:50 couplers; blue oval: wavelength division multiplexer; D: balanced photodetector; dashed black lines: RF signals. (c) Images from Supplementary Video 1. The real-time output includes the calculated times (from system turn-on), the round-trip propagation distance, the closing speed, and link status. (d) Additional experimental setup to synchronize the two sites over a 0-4 km free-space path to a fixed retroreflector and including an in-line 6-pass swept delay line that mimics high closing velocities (see inset).

The system uses a layered approach: TWTFT with a modulated communication channel for picosecond-level time transfer[25] followed by TWTFT with coherent frequency comb pulses for femtosecond-level time transfer. Frequency-comb TWTFT uses linear-optical sampling (LOS) to achieve femtosecond precision (with a 5-nanosecond ambiguity given by the ~200-MHz comb repetition rate). LOS requires the repetition rate of the two pulse trains transmitted across the link to differ by $\Delta f_r \sim 2$ kHz, which leads to inclusion of a third "transfer" comb X at Site A. (See Fig. 1b.) The timing data from the communication channel and frequency-comb transfer are input into synchronization algorithms that, unlike Ref. 13, resolve the 5-ns ambiguity on the comb pulse-by-pulse to generate four calculated timestamps $\{T_{AA}, T_{AB}, T_{BB}, T_{BA}\}$ (see Methods). These four calculated timestamps can be formally interpreted as in conventional two-way time-transfer wherein one signal departs site A at time $T_{AA}$, as recorded at site A, and arrives at site B at time $T_{AB} = T_{AA} + T_{A \to B} - \Delta t_{AB}$, as recorded at site B where $\Delta t_{AB}$ is the time offset between sites and $T_{A \to B}$ is the time-of-flight from A to B. A second signal departs site B at time $T_{BB}$ and arrives at site A at time $T_{BA} = T_{BB} + T_{B \to A} + \Delta t_{AB}$ where $T_{B \to A}$ is the time-of-flight from B to A. From these calculated timestamps, we find the clock time offset as

$$\Delta t_{AB} = \frac{1}{2}[T_{AA} - T_{AB} - T_{BB} + T_{BA}] + \frac{1}{2}[T_{A \to B} - T_{B \to A}] + \Delta T_{cal} \tag{1}$$

where $\Delta T_{cal}$ is an overall transceiver calibration. Clearly, the time offset is incorrect unless the middle term on the right hand side vanishes, i.e. the link is reciprocal, or unless this term is calculated – in our case to the sub-femtosecond level - and removed.

Velocity-Dependent Reciprocity Breakdown and Systematic Doppler Effects

For the case realized here experimentally, the two clocks are connected via a retroreflector moving at closing velocity $V/2$ away from the clocks. The retroreflector is at a distance $L_A(t)$ from site A and $L_B(t)$ from site B. This scenario mimics time-transfer via a moving, intermediate clock site -- the solution presented here could be generalized to the alternate scenario of a stationary clock A and moving clock B with inclusion of the time dilation effect and choice of reference frame. The initial consequence of motion is the breakdown in reciprocity

$$T_{A \to B} - T_{B \to A} = (V/c)(T_{AB} - T_{BA} + \Delta t_{AB}) + (V/c^2)(L_A - L_B), \qquad (2)$$

to first order in $V/c$, where $c$ is the speed of light. The first term results from the link distance being asynchronously sampled by the pulses traveling each direction, and the second from the finite speed of light, which can be derived from geometric considerations or more formally via Lorentz transformations. At a modest $V=30$ m/s, 4-km link, and 0.5-ms asynchronous sampling ($T_{AB} \neq T_{BA}$), the two terms in (2) yield a non-reciprocal time-of-flight of 50 ps and 1.3 ps, respectively. We include this non-reciprocity correction in (1) to <100 as precision by using the available O-TWTFT data to calculate the asynchronous sampling to sub-femtosecond precision and the speed to 20 µm/s precision at 1-second averaging time. The speed is found from the rate-of-change of the measured time-of-flight (calculated via a different combination of timestamps) over three consecutive measurements.

There are two additional effects of motion which lead to strong systematic timing shifts. Both are consequences of the Doppler shifts, which are large ($10^{-7}$, or 20 MHz, at $V=30$ m/s) and changing as $V$ is not constant. First, the calibration term can no longer be treated as a single overall time delay but instead contains a velocity-dependent component. (See Methods.) Second, the Doppler shifts can couple with the system dispersion to cause picosecond-level timing errors in the calculated timestamps. To avoid this, we calculate the ambiguity function[26] of the heterodyne signal between the incoming and local comb light, and find its peak in real-time (<300 µsec) to <100 as precision by use of a Fourier transform algorithm and the Nelder-Mead search algorithm.

The final synchronization algorithms are implemented in a digital signal processing platform to generate an estimate of $\Delta t_{AB}$ in real time at a 2-kHz measurement rate. Under strong turbulence, signal fades block the exchange of comb pulses and the communication link, but these fades are usually of short duration. Therefore, a Kalman filter allows continuous operation through such dropouts[27]. The filter's output adjusts the phase of the clock at site B to synchronize the clocks at 10-100 Hz feedback bandwidth[27].

Results of time synchronization with quadcopter and swept delay line

Figure 2 shows time synchronization between the two sites A and B over a link that includes both 4 km of turbulent air and the swept delay line operated at closing velocities from 0 m/s to ±24 m/s. The time-of-flight, closing velocity, and calculated time offset are all returned from the O-TWTFT signals. In parallel, the clocks' time offset (i.e. arrival time of labelled optical comb pulses) is measured by the out-of-loop verification. When actively synchronized, the clock times agree with a standard deviation of 1.1 fs at the full 2.2 kHz update rate. During brief signal fades due to atmospheric turbulence, the clocks' times walk off randomly (cyan trace of Figure 2(b)) but are resynchronized when the signal is re-acquired.

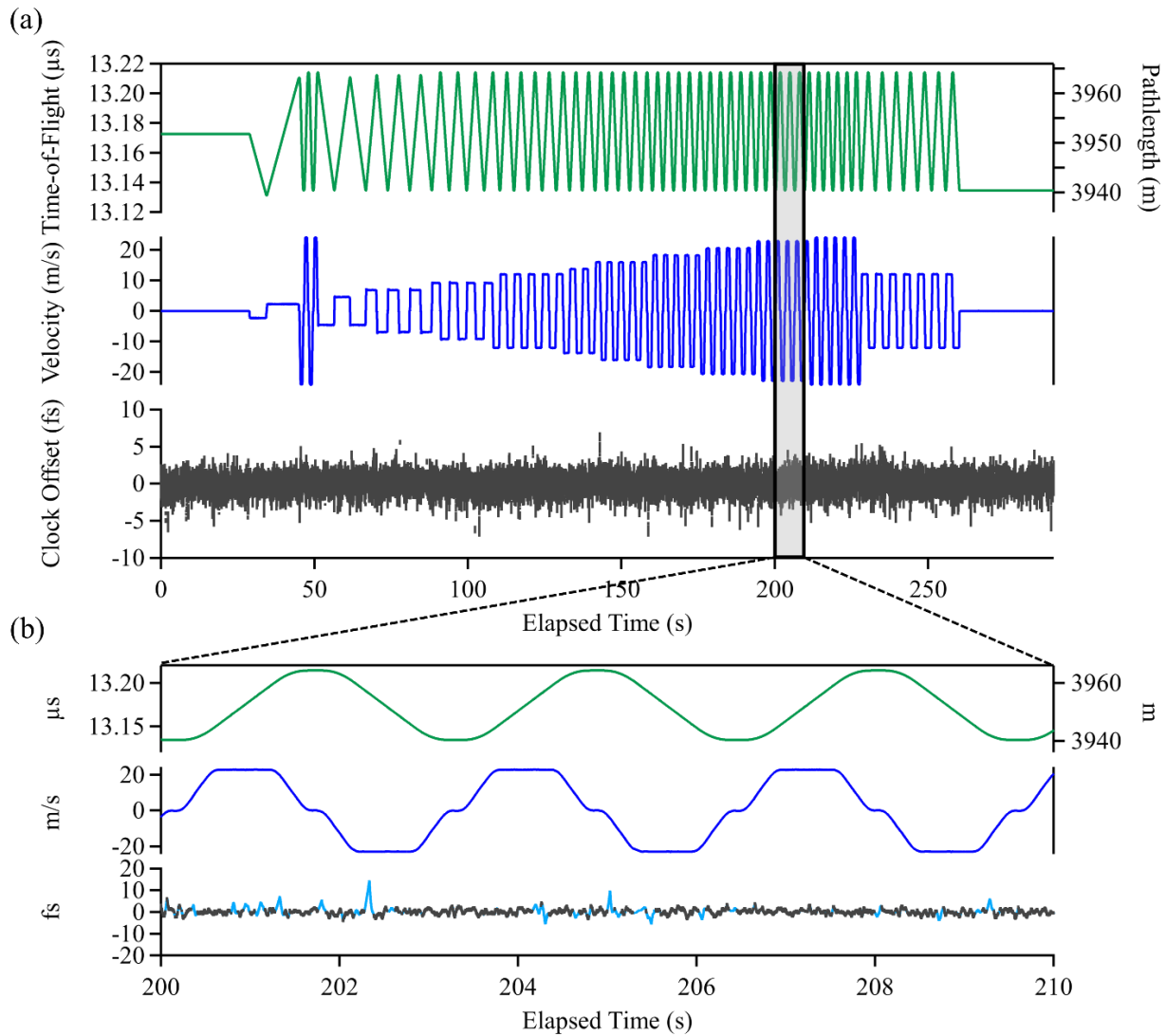

Figure 2: (a) Synchronization over 4 km with the in-line swept delay line operated at closing velocities ranging from 0 m/s to 24 m/s. The time-of-flight (top panel, left axis) and closing velocity (middle panel) are retrieved from the O-TWTFT data. The clock time offset (bottom panel) is the out-of-loop verification. During active synchronization (i.e. no long fades) the standard deviation is 1.1 fs. All data is at the 2.2 kHz update rate. (b) Expanded view. The clocks' time offset is shown for all time (cyan) and only during active synchronization, i.e. no turbulence-induced fades (black line).

Synchronization to a quadcopter-mounted retroreflector is shown in Supplementary Video 1 and Figure 3. The quadcopter provided a maximum 500-meter optical pathlength change and a 20 m/s (quadcopter-limited) maximum speed. Again, we see femtosecond-level synchronization with no evidence of speed-dependent bias. These data do show much longer fades due to the additional challenge of tracking the moving quadcopter[28].

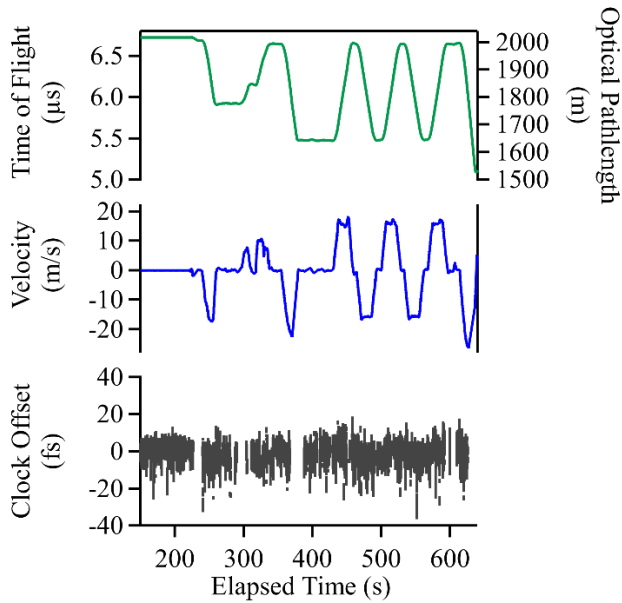

Figure 3: Synchronization results for a link to the flying quadcopter showing the pathlength (top panel), closing velocity (middle panel), and clocks' time offset (bottom panel), measured by the out-of-loop verification channel during periods without signal fades. The standard deviation is 3.7 fs at the ~2 kHz update rate. (Also see Supplemental Video 1.)

Analysis of time and frequency precision (Allan deviation)

Figure 4 shows the time and modified Allan deviations. For these data, the swept delay line was operated for ~20 minutes at ±24 m/s with free-space links of 0, 2, and 4 km. The resulting time deviations, calculated from the out-of-loop verification, all remain below 1 fs for averaging times from 0.1 seconds (the inverse of the synchronization bandwidth) to 200 seconds and are essentially unchanged from a static 0-km shorted measurement. For the quadcopter, the time deviation remains at ~1 fs, elevated above the delay-line data due to longer fades and calibration uncertainties associated with the tracking terminal. The relative fractional frequency instability (modified Allan deviation) for the swept-delay line data is below $10^{-15}$ at a 1-second averaging and $10^{-18}$ at 200-second averaging for all closing velocities. For the quadcopter data, it is $2\times10^{-15}$ at a 1-second averaging and $2\times10^{-17}$ at a 100-second averaging.

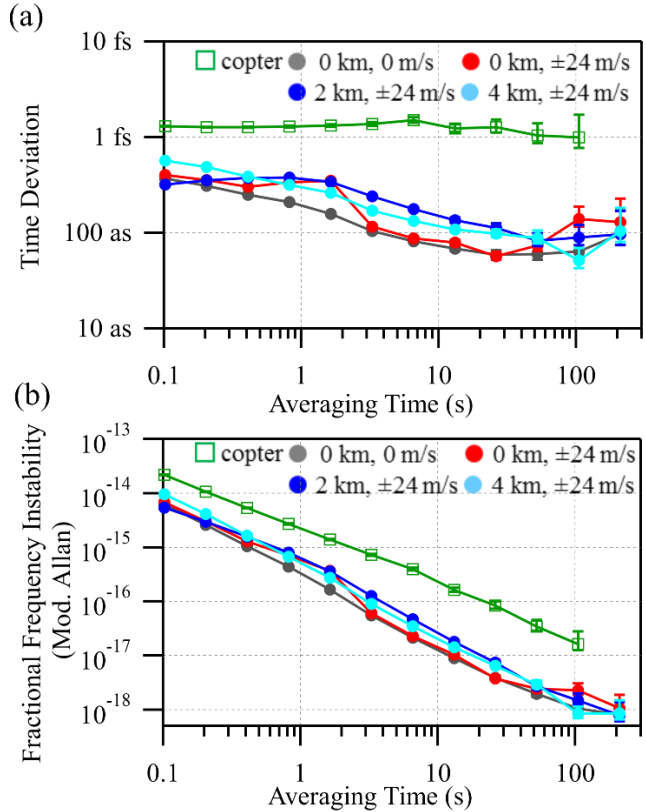

Figure 4: (a) Time deviation for synchronization off the quadcopter with 0-20 m/s motion (open green squares) and with the in-line swept delay line with ±24 m/s motion and a free-space path length of 0 m (red circles), 2 km (blue circles), and 4 km (cyan circles). Also shown is the time deviation at 0 m/s and 0 m free-space path (black circles). The elevated time deviation for the quadcopter data is due to longer signal dropouts and calibration uncertainties associated with the tracking terminal. The O-TWTFT synchronization bandwidth was 10 Hz. (b) Corresponding modified Allan deviation.

**Discussion**

This same approach should scale to the far greater closing velocities of future airborne and satellite-borne clocks although this remains to be tested. Additionally, ground-to-satellite links will suffer from non-reciprocity due to "point-ahead" effects, but theory and ground-based experiments indicate minimal impact[29–31]. The advanced comb-based O-TWTFT described here thus should enable future networks with sub-femtosecond synchronization and $10^{-18}$ frequency syntonization to support multiple time-frequency based precision measurements.

**Methods**

The advanced comb-based O-TWTFT demonstrated here differs significantly in hardware, calibration, and algorithms from the previous comb-based O-TWTFT described in Refs. 13–15. The full extent of the hardware and algorithms will be discussed in a follow-on article[32]. Below we provide a brief outline the hardware and an alternative derivation of a master synchronization equation to the use of the virtual calculated four time stamps of Eq. (1).

Hardware

The overall transceiver structure is similar to Ref. 13, but with major alterations to allow for the presence of motion. Briefly, the overall dispersion in the transceivers was reduced thirty-fold by use of dispersion compensation, the transceiver calibration was performed by a custom, integrated optical time domain reflectometer (OTDR), the rf system was redesigned to reduce multipath reflections due to impedance mismatches, the rf group delay were measured and digitally cancelled prior to interferogram detection, and the digital signal processing hardware was redesigned to support the more extensive, real-time synchronization algorithms. A more detailed description of the redesigned transceiver is given in Ref. 32.

Synchronization algorithm

The presence of motion in the O-TWTFT link required entirely new synchronization algorithms and implementation. The derivation of the four effective timestamps is lengthy and provided in Ref. 32. We briefly outline a different derivation here that leads to a single master synchronization equation but does not provide the same physical insight.

At site A, the comb pulses arrive at the local defined reference plane at times $t_A = n_A f_r^{-1} + \tau_A$, where the integer $n_A$ labels the pulses, $f_r$ is the nominal defined repetition frequency and $\tau_A$ is the overall time offset that includes any integrated frequency error in the local comb. At site B, the comb B pulses arrive at the local defined reference plane at $t_B = n_B f_r^{-1} + \tau_B$, where again $\tau_B$ is the slowly varying clock offset. For verification purposes, we locate the reference plane for both sites at the end of the optical fiber that is used for the out-of-loop time synchronization measurement so that this out-of-loop measurement directly yields $\Delta t_{AB} \equiv t_A - t_B$.

The linear optical sampling (LOS) detection used in comb-based O-TWTFT requires the pulse trains of the two combs transmitted across the link have repetition rates differing by $\Delta f_r \sim 2$ kHz[13]. Therefore, we introduce a third, transfer comb X at the master site with repetition rate $f_r + \Delta f_r$ and time offset $\tau_X$, also phase-locked to the master optical oscillator.

We measure three heterodyne signals between the master, transfer, and remote combs, each consisting of a series of consecutive interferograms, i.e. short heterodyne pulse envelopes, as the pulses cross each other at a rate $\Delta f_r$. If we label each interferogram by an integer $p$ with appropriate subscripts, the three signals are:

$$I_{AX}(t) \approx \sum_{p_{AX}} I_{AX}(t - t_{AX}[p_{AX}]) \quad : \text{Master} \rightarrow \text{Transfer},$$
$$I_{BX}(t) \approx \sum_{p_{BX}} I_{BX}^V(t - t_{BX}[p_{BX}]) \quad : \text{Remote} \rightarrow \text{Transfer}$$
$$I_{XB}(t) \approx \sum_{p_{XB}} I_{XB}^V(t - t_{XB}[p_{XB}]) \quad : \text{Transfer} \rightarrow \text{Remote}$$

where $I_{BX}^V(t)$, $I_{XB}^V(t)$, and $I_{AX}(t)$ are the interferogram pulse shapes, and $t$ is some oracle timescale. The first interferogram, $I_{AX}$, is generated from the local heterodyne mixing of the master and transfer comb at the master site. The middle interferogram, $I_{BX}$, is the heterodyne signal between the transmitted remote comb pulses and the transfer comb at the master site. The third interferogram, $I_{XB}$, is the heterodyne signal between the transmitted transfer comb and the remote comb at the remote site. We are ultimately interested in the times $t_{AX}[p_{AX}], t_{BX}[p_{BX}]$, and

$t_{XB}[p_{XB}]$ that define the centers of the successive interferograms since we will combine these timing data with the communication-based two-way time-frequency transfer to evaluate $\Delta t_{AB}$.

Unfortunately, the interferograms' waveform depends on the Doppler shift of the incoming light (and therefore on velocity as indicated by the superscript) and this, coupled with the relative chirp between the comb pulses, can cause an apparent and strong timing shifts. To avoid this, we calculate the ambiguity function[26] of this heterodyne signal and find its peak in real-time (<300 μsec) by use of a Fourier transform algorithm and the Nelder-Mead search algorithm[32]. With this approach, we suppress this systematic to below 100 as.

Assuming successful suppression of this velocity-induced bias and following similar analysis as in Ref. 14, the three times are

$$
\begin{aligned}
t_{pAX} &= \Delta f_r^{-1} \left\{ p_{AX} - [f_r + \Delta f_r]\tau_X + f_r \tau_A \right\} \\
t_{pBX} &= \Delta f_r^{-1} \left\{ p_{BX} - f_r T_{B\to A}(t_{pBX}) + f_r \tau_B - [f_r + \Delta f_r]\tau_X \right\} \\
t_{pXB} &= \Delta f_r^{-1} \left\{ p_{XB} + [f_r + \Delta f_r]T_{A\to B}(t_{pXB}) + f_r \tau_B - [f_r + \Delta f_r]\tau_X \right\}
\end{aligned}
\quad (3)
$$

as measured with respect to the oracle timescale $t$. In the system however, the timestamps are instead measured against the local timescale at site A or B. Therefore, we use the relationships $t_{pAX} = f_r^{-1} k_{pAX} - \tau_A$, $t_{pBX} = f_r^{-1} k_{pBX} - \tau_A$, and $t_{pXB} = f_r^{-1} k_{pXB} - \tau_B$, where $\tau_{A(B)}$ are the time offsets of the site A(B) timescale from oracle time and $f_r^{-1} k$ is the definition of a local time with $k$ representing the not-necessarily integer ADC sample number. Note that the function $T_{A\to B}(t)$ is the time-of-flight for a signal that arrives at B at time $t$, as measured in oracle time. To solve these equations for the time offset between sites, $\Delta t_{AB} = \tau_A - \tau_B$, we need the integer values, $p_{AX}$, $p_{BX}$

, and $p_{XB}$, as extracted from the communication-based O-TWTFT. We also need the time-of-flight non-reciprocity, Eq. (2), which is briefly derived here.

From (3) and the substitution mentioned afterwards, it is clear we are interested in the asymmetry $T_{A \to B}\left(f_r^{-1} k_{pXB} - \tau_B\right) - T_{B \to A}\left(f_r^{-1} k_{pBX} - \tau_A\right)$. To first order in $V/c$,

$$T_{A \to B}\left(f_r^{-1} k_{pXB} - \tau_B\right) = T_{A \to B}\left(f_r^{-1} k_{pBX} - \tau_A\right) + (V/c)\left[f_r^{-1} k_{pXB} - f_r^{-1} k_{pBX} + \Delta t_{AB}\right], \quad (4)$$

where the second term is the asynchronous sampling contribution. Because of the finite speed of light, $T_{A \to B}(t) - T_{B \to A}(t) = (V/c)\left[L_A(t) - L_B(t)\right] + O\left((V/c)^2\right)$. Combined with (4), this yields the breakdown in reciprocity,

$$T_{A \to B}\left(f_r^{-1} k_{pXB} - \tau_B\right) - T_{B \to A}\left(f_r^{-1} k_{pBX} - \tau_A\right) = (V/c)\left[f_r^{-1} k_{pXB} - f_r^{-1} k_{pBX} + \Delta t_{AB}\right] \\ + (V/c^2)\left[L_A - L_B\right] \quad (5)$$

With this information, we can solve to find the time offset,

$$\Delta t_{AB} = \frac{1}{2 - V/c + \Delta f_r / f_r} \left\{ \frac{\Delta f_r}{f_r^2}\left\{2k_{pAX} - k_{pXB} - k_{pBX}\right\} + \frac{\Delta f_r}{f_r} T_{A \to B}\left(t_{pXB}\right) \right. \\ + f_r^{-1}\left[p_{XB} + p_{BX} - 2p_{AX}\right] + 2\Delta T_{cal} \\ \left. + \frac{V}{c}\left(f_r^{-1} k_{pXB} - f_r^{-1} k_{pBX} + c^{-1}\left[L_A - L_B\right] + 2\Delta T_{cal}^V\right) \right\} \quad (6)$$

where we introduce two calibration terms, $\Delta T_{cal}$, $\Delta T_{cal}^V$, discussed below, and drop terms of order $(V/c)^2$ and higher throughout. In the real-time computation, the closing velocity, $V$, is found by a combination of centered numerical derivatives using the previous three measurements of $T_{A \to B}$ and $T_{B \to A}$, which assumes constant acceleration over $3/\Delta f_r \sim 1.5$ ms.

Transceiver calibration

The calibration term, $\Delta T_{cal}$, nominally reflects a time delay in the transceiver between the reference plane and the incoming pulse detection. However, in reality, each transceiver consists of multiple optical and rf paths between, for example, the optical oscillator, the frequency combs, the optical detection of the arriving frequency comb pulses, the various analog-to-digital converters, and throughout the communication-based O-TWTFT. Without motion, all these paths can be lumped into a single overall time delay. With motion and the resulting Doppler shifts, some delay paths must be corrected for velocity. As a result, the overall transceiver must be calibrated via a built-in rf-domain optical time domain reflectometer (OTDR) that measures the various required delays[32]. In a simplified view, the net result is that the calibration becomes velocity-dependent as

$$\Delta T_{cal} \to \Delta T_{cal} + (V/c)\Delta T_{cal}^{V}.$$

**Acknowledgements**

This work was funded by the National Institute of Standards and Technology (NIST) and the Defense Advanced Research Projects Agency (DARPA) PULSE program. We thank Prem Kumar, Martha Bodine, Jennifer Ellis and Kyle Beloy for helpful discussions.

**Author Contributions**

HB, LCS, JDD, and NRN determined the effects of motion and designed the algorithms. HB, and JDD implemented the digital signal processing. HB, JDD, IK, LCS, and WCS acquired and analyzed the data. KCC designed and implemented the tracking terminal necessary for quadcopter operation. WCS designed and implemented the swept delay line. MC and KCC assisted with hardware implementation and quadcopter operation. LCS, JDD, and NRN prepared the manuscript.

**Competing Interests**

The authors declare no competing financial interests.